\documentclass[pra, notitlepage, superscriptaddress,  reprint, twocolumn,nofootinbib]{revtex4-1}
\usepackage{comment}
\usepackage{enumerate}
\usepackage{amssymb}
\usepackage{amsmath}
\usepackage{braket}
\usepackage{graphicx}
\usepackage[usenames,dvipsnames]{color}
\usepackage{tensor}
\usepackage{empheq}
\usepackage{capt-of}
\usepackage[normalem]{ulem} 
\usepackage{soul} %
\usepackage{MnSymbol}
\usepackage{leftidx}
\usepackage{tcolorbox}
\usepackage{dsfont}
\usepackage[thinlines]{easytable}
\usepackage{mdframed}

\usepackage[colorlinks,bookmarks=false,citecolor=NavyBlue,linkcolor=OliveGreen,urlcolor=blue]{hyperref}
\newcommand{\be}{\begin{equation}}
\newcommand{\ee}{\end{equation}}
\newcommand{\ba}{\begin{aligned}}
\newcommand{\ea}{\end{aligned}}
\newcommand{\bw}{\begin{widetext}}
\newcommand{\ew}{\end{widetext}}

\newcommand{\bea}{\begin{eqnarray}}
\newcommand{\eea}{\end{eqnarray}}

\def\doi{http://dx.doi.org/}

\renewcommand{\bra}[1]{\left< #1 \right|}
\renewcommand{\ket}[1]{\left| #1 \right>}
\newcommand{\sandwich}[2]{\left<\left.#1\ \right| #2\, \right>}
\newcommand{\sqs}{\text{\sc sqs}}
\newcommand{\D}{\rm d}

\definecolor{gruen}{rgb}{0,0.625,0}

\begin{document}
\title{In and out of equilibrium quantum metrology with mean-field quantum criticality}
\author{Sascha Wald}
\email{swald@pks.mpg.de}
\affiliation{
Max-Planck-Institut f\"ur Physik Komplexer Systeme, N\"othnitzer Stra{\ss}e 38, D-01187, Dresden, Germany\looseness=-1
}
\author{Saulo V. Moreira}
\affiliation{Centro de Ci\^encias Naturais e Humanas, Universidade Federal do ABC - UFABC, Santo Andr\'e, Brazil
}
\author{Fernando L. Semi\~ao}
\affiliation{Centro de Ci\^encias Naturais e Humanas, Universidade Federal do ABC - UFABC, Santo Andr\'e, Brazil
}

\begin{abstract}
We study the influence that collective transition phenomena have on quantum metrological protocols. The single spherical 
quantum spin (SQS) serves as stereotypical toy model that allows analytical insights on a mean-field level.
First,  we focus on equilibrium quantum criticality in the SQS and obtain the quantum Fisher information analytically, which is associated with the minimum lower bound for the precision of estimation of the parameter driving the phase transition.
We compare it with the Fisher information for a specific experimental scenario where photoncounting-like measurements are employed. We find how quantum criticality and squeezing are useful resources in the metrological scenario.
Second, we obtain the quantum Fisher information for the  out of equilibrium transition in the dissipative 
non-equilibrium steady state (NESS), and  investigate how the presence of dissipation affects the parameter estimation. In this scenario,
 it is known that the critical point is shifted by an amount which depends on the dissipation rate. This is used here to design high precision protocols for a whole range of the transition-driving parameter in the ordered phase. In fact, for certain values of the parameter being estimated, dissipation may be used to obtain higher precision when compared to the equilibrium scenario.

\end{abstract}

\maketitle

\newpage
\section{Introduction}

Bridging the gap between fields such as quantum optics, quantum information and 
statistical mechanics is a challenging but fruitful research direction. 
Interdisciplinary knowledge transfer allows these fields to mutually benefit from one another and, it is often the case that certain techniques or concepts initially tailored for a certain problem, say in statistical mechanics, yield a distinct perspective on and a deeper understanding of challenges in quantum information or optics.
%
Perhaps the most prominent example is the study of phase transitions from the perspective of quantum information.  Phase transitions and bipartite entanglement, for instance, are known to be closely related at arbitrary temperatures \cite{Wei2018}. Scaling of entanglement near to a quantum critical point has been studied in different systems \cite{Fazio2002,Kitaev2003}. Other quantum information tools such as distance measures and state fidelities \cite{Zhou, Zanardi2, Zanardi3, Cozzini, Cozzini2, Bina2,Yang2008}, state coherence \cite{Lin2016}, and general quantum correlations \cite{Amico2012,Rigolin2013} have also been extensively used in the study of quantum phase transitions.

In general, physical parameters such as coupling constants cannot be directly measured, and therefore need to be estimated in an indirect way through observables or generalized measurements.
Quantum metrology is concerned with the estimation
of such quantities by exploiting quantum resources in order to improve the precision of the estimation protocol.
For example, quantum features such as entanglement \cite{Toth},
squeezing \cite{Maccone}, quantum temporal correlations \cite{Moreira,Mor17}, and quantum invasiveness \cite{Mor19}
can be associated with a significant improvement of sensitivity in metrological protocols.
If the global state of a system under inspection 
is highly sensitive to small variations of a certain parameter, this may be used to increase the precision with which this parameter can be estimated \cite{Yuan,Zanardi,Giovannetti}. This provides a route to investigate critical systems from the point of view of quantum metrology, as we are going to pursuit in this work \cite{Frerot, Gar19, Zanardi, Carollo, Braun}.

Here, we focus on a quantum mean-field version of the celebrated spherical model \cite{Berl52,Lew52,Henk84,Voj96}
which is a useful tool in statistical mechanics for the study of strongly interacting
degrees of freedom. 
Based on the idea of introducing transition phenomena through adequate constraints or appropriate limits of physical parameters in few body problems \cite{Ashhab2013,Plenio2015}, we study here the case of a single SQS \cite{Wald16}.  From the point of view of statistical mechanics, the SQS may be interpreted as a mean-field version of the spherical model, while from 
the point of view of quantum optics and quantum thermodynamics, it may be seen as a highly controlled quantum system 
subjected to work protocols, as explained in \cite{Timpa19}. Originally introduced to overcome the lack of an 
analytical solution of the three dimensional Ising model \cite{Berl52}, the spherical model has 
quickly proven itself as an excellent starting ground to study various transition
phenomena in and out of equilibrium and routinely obtain results that go beyond 
mean-field statements \cite{Singh81,Voj96,Wald15,Wald18}.

The SQS may be viewed as one of the simplest quantum systems that still 
allows for transition phenomena to take place. In thermal equilibrium,
the SQS has a quantum critical point separating a paramagnetic (disordered) from a ferromagnetic (ordered) phase.
Therefore, states close to this critical value do
show macroscopically different properties due to large fluctuation 
effects at quantum criticality. Furthermore, the transition phenomenon is stable against thermal perturbations, meaning that the SQS shows a continuous 
critical line for $T>0$. We shall therefore propose a metrological protocol for the SQS.
Typically, a metrological protocol is divided into four different stages \cite{Esc11, Esch11}:
{\it (i) preparation}: a certain system state is initialized, {\it (ii) sensing}: a parameter is imprinted in the 
system's state via a certain dynamics, {\it (iii) readout}: the system is measured,
 {\it (iv) estimation}: the parameter is estimated from the measurement outcomes.

In our case, however, the sensing stage is to be seen as part of the preparation of a steady state which depends on the parameter driving the phase transition.
We intend to estimate this parameter in two distinct steady state scenarios:
First, in a $T=0$ thermal equilibrium state and second in a specific NESS.
The use of NESS transitions in quantum metrology \cite{Banchi, Carollo2, Marzolino} is rather unexplored in the literature 
compared to its equilibrium counterpart \cite{Bina,Zanardi}.
Previous studies have shown examples illustrating the usefulness of driven and dissipative dynamics \cite{Zuppardo} and NESS transitions \cite{Gar19} for metrology.
Here, we provide an extension to the non-equilibrium scenario by exploiting the metrological potential of 
NESS transitions in the SQS. This study helps to highlight the potential of NESS compared 
to equilibrium states as both cases can be worked out and compared explicitly in the SQS.

This paper is organized as follows.
In section~\ref{sec:sqs} we analyze the equilibrium quantum metrology. 
Therefore we first briefly introduce the Fisher information and the quantum
Fisher information as central metrological quantities. We then
discuss
the thermal equilibrium of the SQS in 
the language of quantum optics. Next, we discuss how the zero temperature quantum phase 
transition may be used for a precision gain in parameter estimation protocols.
In section~\ref{sec:outofeq} we introduce the driven SQS and review its 
NESS transition, which we shall then exploit as a quantum metrological resource.
Finally we discuss how such NESS transitions may be beneficial compared to
equilibrium quantum metrology.



\section{SQS Equilibrium Quantum Metrology}
\label{sec:sqs}

In the present work, we shall put forth a metrological study of the 
SQS \cite{Wald16} by evaluating the quantum Fisher information and the Fisher information associated with a specific measurement scenario.
Before moving on, we briefly review the meaning of these quantities in metrological protocols.

The Fisher information is a figure of merit for a metrological protocol concerned with the estimation of a certain parameter $g$ encoded in the physical state of the system. 
It gives the sensitivity of the estimation as it is connected to the lower bound of the standard deviation $ \Delta g $,
\begin{equation}
\Delta g \ge \frac{1}{\sqrt{\nu F(g)}}   \ ,
\end{equation}
where $\nu$ is the number of realizations of the experiment, for unbiased measurements \cite{Fisher, Paris}. 
Therefore, the Fisher information depends on the specific measurement that one performs on the system.
If we term the probabilities associated with each possible result $l$ of the measurement as $p_l(g)$, satisfying $\sum_l p_l(g)=1$,
then the Fisher information may be written as
\begin{equation}\label{eq2}
F(g)=\sum_l p_l(g)\left[\frac{\partial \ln p_l(g)}{\partial g}\right]^2.
\end{equation}
By expressing the probabilities as $p_l(g) = {\rm Tr}(\rho(g)E_l)$ where $\rho(g)$ depends on $g$ and $\{E_l\}$ is a \textit{positive operator valued measure} (POVM), the Fisher information is generalized to quantum mechanics.
The upper bound for $F(g)$ is called the quantum Fisher information $\mathcal{F}$   \cite{Helstrom, Holevo, Braunstein, Braunstein2}, and corresponds to the maximization of  $F(g)$ over all quantum measurements $\{E_l\}$,
\begin{equation}
\mathcal{F} = \max_{\{E_l\}}F(\rho,\{E_l\}) .
\end{equation} 
Therefore, the quantum Fisher information is the Fisher information associated with the optimal measurement, i.e. the one associated with the maximal precision for the estimation of $g$.

\subsection{The SQS in Thermal Equilibrium}

The SQS is a simple toy model that was designed
to study the effects of collaborative transition phenomena in a rather easy 
quantum optics language.
As such, it is described by a single and coherently driven quantum harmonic 
oscillator with the Hamiltonian
\begin{subequations}
\begin{align}
    \label{eq:H}
 H &= \frac{g}{2}p^2 + \frac{\omega^2}{2g}x^2 - h_0 x,
 \end{align}
 that is subject to the following two distinct constraints \cite{Wald16,Timpa19}
 \begin{align}
 \label{eq:constraints}
 \left<x^2\right> &= 1, \quad  h_0 = \left< x \right>.
\end{align}
\end{subequations}
Here, $x$ and $p$ are canonically conjugate variables describing the position and 
momentum of the oscillator and satisfying the standard bosonic commutation relation 
$[x,p] = i$, $\omega$ is the frequency of the oscillator, $h_0$ parametrizes the coherent driving and $g$ is the (inverse) mass of the quantum oscillator.
The constraints that define the SQS are formulated on statistical averages
$\left< O \right> := \operatorname{tr}(\rho O) $ with $\rho$ being the 
density matrix of the system.

The SQS can be derived as a molecular-field approximation of the full $N$ body
quantum spherical model \cite{Voj96,Henk84} and is thus a mean-field version of the latter \cite{Wald16}. In this context
the parameter $g$ is usually referred to as {\it quantum parameter} since the case $g=0$ 
corresponds to the classical spherical model \cite{Berl52}.  We shall adopt this 
terminology and therefore refer to the mass as quantum parameter.
The first constraint in Eq. (\ref{eq:constraints}) is the one-body version 
of the $N$-body spherical constraint \cite{Berl52} and confines fluctuations in the 
position degree of freedom. 
The second constraint in Eq.~(\ref{eq:constraints}) can be seen as the aforementioned
molecular field approximation. Here, a complicated many-body interaction is replaced 
by a self interaction of a single degree of freedom, as it is routinely done for
the Weiss theory of magnetism \cite{Yeo92}, for example. 
Together, these constraints render the system effectively non-linear and therefore 
allow for different physical phases to exist. We shall review the 
order-disorder transition of this system in the following.

We first introduce bosonic ladder
operators $(a,a^\dagger)$ obeying the bosonic commutation relation $[a,a^\dagger] = 1$,
\begin{align}
 x &= \sqrt{\frac{g}{2\omega}} \left(a+a^\dagger \right),\qquad  
 p = i\sqrt{\frac{\omega}{2g}} \left(a^\dagger -a\right).
\end{align}
In terms of these creation and annihilation operators, the Hamiltonian reads
\begin{align}
    \label{eq:Ha}
 H &= \omega \left(a^\dagger a + \frac{1}{2}\right)- h \left(a+a^\dagger \right),
\end{align}
with the rescaled driving given by $h = h_0 \sqrt{g/2\omega}$. 
In order to diagonalize the Hamiltonian, we need to introduce displaced operators
\begin{equation}\label{eq:shift}
 b = a - \alpha = \mathcal{D}(\alpha) a \mathcal{D}^\dagger(\alpha),
\end{equation}
where  $\mathcal{D}(\alpha) = \exp\left(\alpha a^\dagger - \alpha^\ast a \right)$
is the displacement operator.
The displaced vacuum is $\ket{0_b} = \mathcal{D}(\alpha) \ket{0_a}$, which corresponds to
the ground state of the transformed Hamiltonian
\begin{equation}\label{eq:Hb}
 \tilde{H }= \omega \left(b^\dagger b +\frac{1}{2} \right) -\frac{h^2}{\omega},
\end{equation}
obtained by choosing $\alpha= h/\omega$. In order to study the equilibrium
properties of the system we first imagine it being weakly coupled to a thermal
heat bath that eventually thermalizes the system. Such a process may be described 
by the following standard relaxational Lindblad
master equation for the reduced density matrix $\rho$
\begin{align}
 \begin{split}
 \partial_t \varrho =  -i \left[\tilde{ H},\varrho\right] &+ \gamma (n+1) \left( 
 b \varrho b^\dagger - \frac{1}{2} \left\{ b^\dagger b , \varrho \right\}\right) \\
 &+\gamma n \left( 
  b^\dagger\varrho b - \frac{1}{2} \left\{b b^\dagger  , \varrho \right\} \right),
   \end{split}
\end{align}
where $n=\frac{1}{2}\coth\left( \omega/2T \right) - \frac{1}{2}$ is the Bose-Einstein distribution.
In the steady state $\rho_{ss}$, which corresponds to the equilibrium state, 
the ladder operators obey the following set of expectation values
\begin{align}
 \left<b\right> &= 0, \quad \left< bb \right> = 0,\quad
 \left<b^\dagger b\right> = \frac{1}{2}\coth(\omega/2T) -\frac{1}{2}.
\end{align}
In the non-displaced picture these expectation values read
\begin{align}
 \left< a \right>= \frac{h}{\omega}, \hspace{.25cm} \left< aa \right> = \frac{h^2}{\omega^2}, \hspace{.25cm}
 \left<a^\dagger a \right> = \frac{\coth \left( \frac{\omega}{2T}\right)-1}{2} + \frac{h^2}{\omega^2},
\end{align}
and allow us to determine straightforwardly the self-consistent parameters 
$\omega$ and $h$ from the external constraints. These allow for two distinct
solutions viz \cite{Timpa19}
\begin{align}
  \omega &= \frac{g}{2}\coth \left( \frac{\omega}{2T}\right), \qquad h = 0.
 \label{eq:disorder}
 \\[.5cm]
  \omega &= \sqrt{g}, \qquad  h  =\left[ \frac{\sqrt{g}}{2} -\frac{g}{4}\coth \left( \frac{\sqrt{g}}{2T}\right)\right]^{1/2}.
  \label{eq:order}
\end{align}
Eq.~(\ref{eq:disorder}) is called the {\it disorder solution} and
Eq.~(\ref{eq:order}) is called the {\it ordered solution}.
This nomenclature comes from the fact that the self-consistent 
magnetic field is proportional 
to the parameter $\alpha := \left< a \right>$. This parameter quantifies the displacement of the oscillator in
its equilibrium position, and therefore is an {\it order parameter } for the system. It is interesting to observe that 
the shift in Eq.~(\ref{eq:shift}) is only needed to diagonalize the Hamiltonian in the ordered phase.

\begin{figure}[t!]
\centering
\includegraphics[width=.45\textwidth]{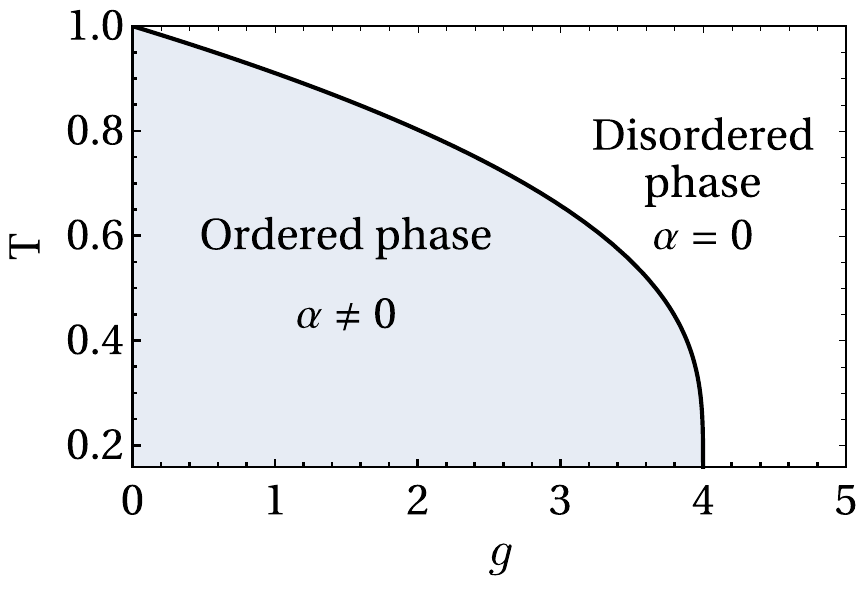}
 \caption{Equilibrium phase diagram of the SQS. Two distinct phases 
 are present, namely a disordered phase with an average magnetization equal to zero
 and an ordered phase with a non-zero average magnetization. The phases are 
 separated by a thermal critical line that ends in a zero temperature quantum 
 phase transition.}
 \label{fig:phases}
\end{figure}

The critical $g$-$T$ line that separates the ordered from the disordered phase is found from comparing both
solutions and, therefore, is described by the functional relation
\begin{equation}
 \frac{2}{\sqrt{g_c}} = \coth\left( \frac{\sqrt{g_c}}{2T_c} \right).
\end{equation}
The equilibrium phase diagram is depicted in Fig.~\ref{fig:phases}.

\subsection{Equilibrium Parameter Estimation}

As mentioned before, we shall be concerned with the estimation of the quantum parameter $g$, which is the parameter driving the quantum phase transition. 
From here on, we shall work at zero temperature $T=0$ since we want to explore the effects of true quantum criticality rather than thermal criticality. 
This is motivated by the 
fact that we want to understand the impact of a genuine quantum transition for metrology
and that the $T>0$ transition is effectively classical.

First, we calculate the quantum Fisher information.
At $T=0$, the system's equilibrium state,  in which the quantum parameter $g$ is imprinted, 
corresponds to the ground state of the SQS, $\rho = \ket{0_{\sqs}(g)}\bra{0_{\sqs}(g)}$.
As this is a pure state, the quantum Fisher information can therefore be expressed as
\begin{align}\label{deff}
\mathcal{F} &= 4\left(  \frac{\D \bra{0_{\text{\sc sqs}}}}{\D g} \frac{\D \ket{0_{\text{\sc sqs}}}}{\D g}   
- \left|\frac{\D \bra{0_{\text{\sc sqs}}}}{\D g} \ket{0_{\text{\sc sqs}}}\right|^2  \right).
\end{align}
According to Eq.~(\ref{deff}), we need to evaluate the derivative of the 
SQS ground-state with respect to the imprinted quantum parameter $g$.
Several ways of how this is achieved can be found in appendix~\ref{app:qfi}.
The quantum Fisher information then reads
\begin{align}\label{quantumFisher}
\mathcal{F} =\left( 2\frac{\D\alpha}{\D g} 
+\frac{\alpha}{g} \frac{ \omega-g\ \frac{\D\omega}{\D g}}{\omega} \right)^2
+ \frac{1}{2}\left(\frac{1}{g}-\frac{\D \log (\omega)}{\D g}  \right)^2.
\end{align}

In order to investigate whether the upper bound set by the quantum Fisher information can be saturated, we evaluate the Fisher information considering 
{\it photoncounting-like measurements}. Such measurements correspond to simple projective measurements $\{ \ket{m_\Omega}\bra{m_\Omega}\}_{m=0,1,...}$, where  
 $\{ \ket{m_\Omega}\}_{m=0,1,...}$ is a Fock basis described by a characteristic frequency $\Omega$.

It is clear from the previous
section that the SQS can always be described by a $g$-dependent
displacement of an underlying harmonic oscillator as
\begin{equation}
\ket{0_\sqs(g)} = \mathcal{D}(\alpha) \ket{0_g}.
\end{equation}
In order to calculate the Fisher information we need to evaluate the probabilities of counting $m$ photons, 
\begin{align}
p_m(g,\Omega) = \left|\sandwich{m_\Omega}{0_{\text{\sc sqs}}}\right|^2 
= \left|\bra{m_\Omega}\mathcal{D}(\alpha)\ket{0_g}\right|^2.
\end{align}

To proceed, it is useful to express the underlying harmonic oscillator ground-state $\ket{0_g}$ in the $\{ \ket{m_\Omega}\}_{m=0,1,...}$ basis. 
Since both of these Fock basis describe harmonic 
oscillators, they may be connected by a Bogoliubov transformation. 
Such Bogoliubov transformations
can be expressed through a squeezing operation which is associated with a
unitary transformation \cite{Bar02,Walls08}.
In this way, we write
\begin{equation}
 a_\Omega = S^\dagger(\zeta) a_g S(\zeta),
\end{equation}
with the squeezing operator 
\begin{equation}
S(\zeta) = \exp\left(\frac{\zeta^\ast}{2}\, a_g^2
- \frac{\zeta}{2}\, \left(a_g^\dagger\right)^2 \right)
\end{equation}
and the
complex squeezing parameter $\zeta(g,\Omega)= r \exp(2i\phi)$.
The single mode squeezed state is then the vacuum of the transformed operators
via $\ket{0_g} = S(\zeta)\ket{0_\Omega}$.

Introducing the squeezed states
\begin{equation}
 \ket{\alpha, \zeta}= \mathcal{D}(\alpha) S(\zeta) \ket{0_\Omega}
\end{equation}
we may write the probabilities as overlap probabilities between the Fock
state and the squeezed state. By means of \cite[Eq.~(3.7.5)]{Bar02} and 
\cite[Eq.~(B.5)]{Cah69} we obtain
 \begin{widetext}
\begin{align}
 p_m(g,\Omega) 
&=\frac{\alpha^{2m}e^{-\left|\alpha\right|^2}\operatorname{sech}(r)}{\Gamma(m+1)\pi^{\frac{1}{4}}}
\left| \sum_{n=0}^\infty\sqrt{\Gamma \left(n+\frac{1}{2}\right)} \left( 
-\frac{e^{i2\phi}}{\alpha}\tanh(r)\right)^n   L_{n}^{(m-n)}(\alpha^2) \right|^2.
\label{eq:proba}
\end{align}
 \end{widetext}
Here $\Gamma(\cdot)$ is the Gamma function and $L_n^{m}(\cdot)$ are the Laguerre polynomials \cite{Abra64}.

In order to evaluate these probabilities we need to determine solely the phase and 
the magnitude of the squeezing parameter $\zeta$. To do this, we explicitly
derive the transformation that maps $a_g$ to $a_\Omega$,
\begin{align}
 \frac{a_g+a_g^\dagger}{\sqrt{2\omega/g}}
 = x = \frac{a_\Omega+a_\Omega^\dagger}{\sqrt{2\Omega}}
 \quad \Rightarrow \quad
 a_\Omega+a_\Omega^\dagger=   \frac{a_g+a_g^\dagger}{\sqrt{\omega/g\Omega }},
 \\
\frac{a_g- a_g^\dagger}{i\sqrt{2g/\omega}} = p =
\frac{ a_\Omega-a_\Omega^\dagger}{i \sqrt{2/\Omega}}
\quad \Rightarrow \quad
a_\Omega^\dagger - a_\Omega = \frac{a_g^\dagger - a_g}{\sqrt{g \Omega/\omega}}.
\end{align}
Consequently, we may write the transformation as
\begin{align}
 a_\Omega &= 
 \cosh (r) a_g - e^{-2 i \phi} \sinh(r) a_g^\dagger,
\end{align}
with the relevant squeezing parameter characterized by
\begin{equation}\label{eq:squeezing_parameters}
 \tanh (r) = \left|\frac{\omega- g \Omega }{\omega+ g \Omega} \right|,
 \quad
 \phi = -\frac{1}{2}\operatorname{arg}\left(\omega- g \Omega  \right).
\end{equation}

\begin{figure*}[t]
 \centering
 \scalebox{.95}{\includegraphics{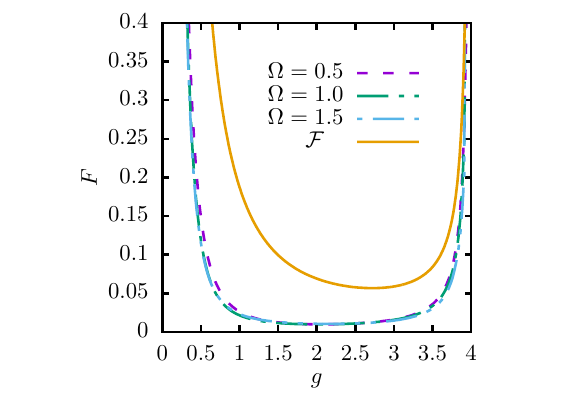} } \hspace{-1.5cm}
 \scalebox{.95}{\includegraphics{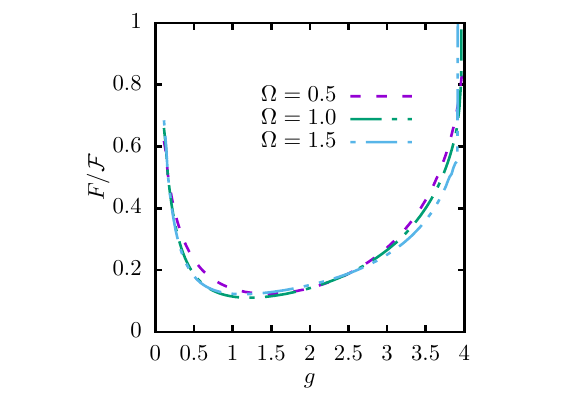} }
 \caption{\underline{Left panel}: The Fisher information $F$ for the SQS system as a function of $g$ for different measurements parametrized by the parameter $\Omega$, and the quantum Fisher information as a function of $g$ as well, which establishes the upper bound for the Fisher information.
 \underline{Right panel}: Normalized Fisher information $F/\mathcal{F}$ as a function of $g$. We can see that photon count measurements tend to be optimal for the estimation of the quantum parameter $g$ for small values of $g$ and as $g$ approaches the critical value $g_c=4$. For measurements associated with smaller values of $\Omega$, however, we see that $F/\mathcal{F}$ can attain larger values in the region close to $g=4$.}\label{plots}
\end{figure*}
%

In Fig.~\ref{plots}, we focus on the ordered phase and plot the Fisher
information as a function of the quantum parameter $g$. In line with previous results for equilibrium states \cite{Gar19,Bina}, we see that the quantum Fisher information diverges at the critical point ($g_c=4$), which means that arbitrarily large precision associated with the estimation of the quantum parameter $g$ could in principle be attained.
Furthermore, the quantum Fisher information diverges as $g\rightarrow 0$ as well,
which implies that at the quantum to classical limit $g\rightarrow 0$, arbitrarily large precision can be attained. 
This means that the ground-state is very sensitive to quantum fluctuations around the classical fix point $g=0$, which is harvested as 
a resource for the quantum Fisher information.

In the right panel of Fig. \ref{plots}, we show the normalized Fisher information $F/\mathcal{F}$ associated with 
the projections $\{ \ket{m_\Omega}\}_{m=0,1,...}$ on the Fock basis,  as a function of $g$ and for different values of the reference frequency $\Omega$.
First, we see that in the interval $0  \lesssim g  \lesssim 0.3$, different projective measurements associated with different Fock basis (i.e. for different values of $\Omega$) give the same $F/\mathcal{F}$. In particular, as $g\rightarrow 0$, $F/\mathcal{F}\rightarrow 1$.
It is also noticeable that projections associated with decreasing values of $\Omega$ can provide a lower standard deviation $\Delta g$ in the range $3  \lesssim g  \lesssim 3.8$, as $F/\mathcal{F}$ increases in this interval as  $\Omega$ decreases.
However, in the range $3.8  \lesssim g < 4$, different $\Omega$ tend to be associated with the same normalized Fisher information, which goes to $1$ as $g$ tends to $4$.

We verified that criticality is a resource for the estimation of $g$, as $F/\mathcal{F}$ is larger in the region close to $g=4$.
This is only valid when the system is in the ordered phase, since the quantum Fisher information is zero in the disordered phase ($g > 4$).
In other words, for any choice of POVM in (\ref{eq2}), the Fisher information will be zero in the disordered phase, meaning that it is not possible to establish a lower bound for precision in the estimation of $g$, i.e. there is no lower bound for the standard deviation $\Delta g$.

Moreover, we shed light on the region $g\ll 1$ where quantum fluctuations around the classical fix point 
yield a strong $g$ dependence of the ground-state properties. This delicate dependence may then be exploited as a metrological
resource.

\begin{figure}[b]
 \centering
   \scalebox{.95}{\includegraphics{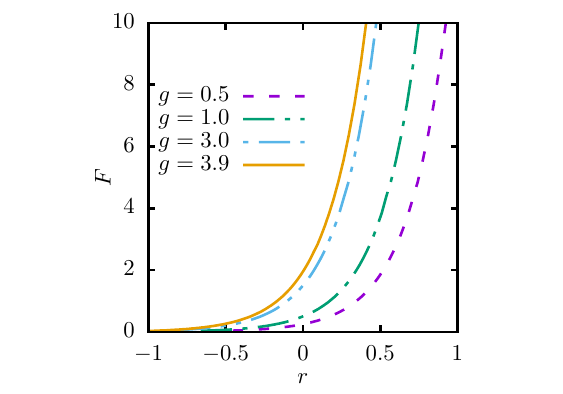}}
  \caption{The Fisher information $F$ for a SQS system parametrized by the fluctuation
 parameter $g$ as a function of the associated magnitude of the squeezing parameter. We can see that the Fisher information increases monotonically as the magnitude of the squeezing parameter increases.}
 \label{fig:squeeze}
\end{figure}

To conclude this analysis, we want to point out that squeezing helps the parameter estimation.
Therefore, in Fig.~(\ref{fig:squeeze}), we show the Fisher information as a function of the magnitude of the squeezing parameter $\zeta$ for fixed values of $g$.
We see that the Fisher information increases monotonically as the magnitude of $\zeta$ increases, which shows that the quality of the estimation of $g$ in the metrological protocol can be improved by increasing the magnitude of the squeezing parameter in quantum mean field systems.
This is in line with previous results which shows that precision benefit from squeezing in metrological protocols \cite{Maccone}.

\section{SQS out of Equilibrium Quantum Metrology}
\label{sec:outofeq}

Having studied the effects of equilibrium criticality on quantum metrology, we now
address the question of whether non-equilibrium state transitions may be beneficial
for the estimation of the parameter $g$.
%
In this section, we first shall review the type of NESS transition in the SQS that we will study as a probe in the metrological protocol.  \cite{Wald16}

\subsection{The NESS transition in the SQS}

\label{ssec:ness}
In this section we shall review the non-equilibrium steady state of the SQS that emerges from the following
Lindblad master equation
\begin{align}
\begin{split}
 \partial_t \varrho =  -i \left[ H,\varrho\right] &+ \gamma (n+1) \left( 
 a \varrho a^\dagger - \frac{1}{2} \left\{ a^\dagger a , \varrho \right\}\right) \\
 &+\gamma n \left( 
  a^\dagger\varrho a - \frac{1}{2} \left\{a a^\dagger  , \varrho \right\} \right).
  \label{eq:NESSL}
  \end{split}
\end{align}
Instead of thermalizing the eigenmode of the coherently driven oscillator, this dynamics thermalizes 
the eigenmodes of the non-driven oscillator which results in a competition between the dissipative dynamics
and the coherent drive.
It is known that this system shows a dissipative NESS transition
\cite{Wald16}. The time evolution governed by Eq.~(\ref{eq:NESSL}) can be engineered by adding controlled degrees of freedom \cite{Zoller93} and by using  external lasers with appropriate intensity and phase fluctuations \cite{Milburn98}. It is well-established that this dynamics relaxes towards a coherent state 
with an amplitude dependent on the driving strength and dissipation rate.

In the NESS, i.e. $ \partial_t \varrho=0$, one finds \cite{Wald16}
\begin{align}
 \left< a \right> &= \frac{1}{2}\sqrt{1-\frac{2\omega}{g}}\left(i -
 \frac{\gamma}{  2\omega}\right), \\
 \left<aa\right>&=\frac{g-2 \omega}{4g\omega}\left[
 \frac{\gamma^2-2 g}{2\omega}
 -i \gamma\right].
\end{align}
These expectation values allow us again to impose the external constraints, as it was done in the equilibrium case. 
It turns out that there are again two distinct solutions, and while the disordered equilibrium solution persists, the 
ordered solution is altered.
The protocols to satisfy the external constraints in the ordered phase are given by \cite{Wald16}
\begin{align}\label{eq:NESSsol}
 \omega &= \frac{1}{2} \sqrt{4 g-\gamma ^2}, \quad 
 h=-\frac{g \gamma }{4 g-\gamma ^2}\sqrt{1-\sqrt{\frac{4 g-\gamma ^2}{g^2}}}.
\end{align}

As in the equilibrium case, these self-consistent protocols determine the displacement parameter $\alpha = h/\omega$. 
In the disordered phase, the displacement is $\alpha = 0$ or equivalently $h=0$. This means that the external force term in Hamiltonian Eq.~(\ref{eq:H}) is null. Therefore, 
the master equation in Eq.~(\ref{eq:NESSL}) is effectively that of an
harmonic oscillator in contact with a thermal bath.
Thus, it is clear 
that the disordered phase is indeed an equilibrium phase while the ordered phase is a non-equilibrium phase.
The critical $g$-$\gamma$ line separating these phases is again found from the condition $h=0$ and it is depicted in Fig.~\ref{fig:NESSPD}.

Note that for certain values of $\gamma$
two distinct phase transitions are observed.
The first, on the right-hand side of Fig. \ref{fig:NESSPD} is ``similar'' to the equilibrium case, as increasing quantum fluctuations parametrized by $g$ destroy macroscopic order.
Nevertheless, this transition is a true non-equilibrium transition.
In turn, the second transition on the left-hand side of Fig. \ref{fig:NESSPD}
is a striking non-equilibrium effect, as increasing quantum fluctuations 
induce macroscopic order \cite{Wald16}.
For the following metrological study, it is important to note that the values of $g$ corresponding to transition points are dependent on the choice of the dissipation rate $\gamma$.

%
%

\subsection{NESS Parameter estimation}
\label{dissipation}
\begin{figure}[t]
\centering
\includegraphics[width = .45\textwidth]{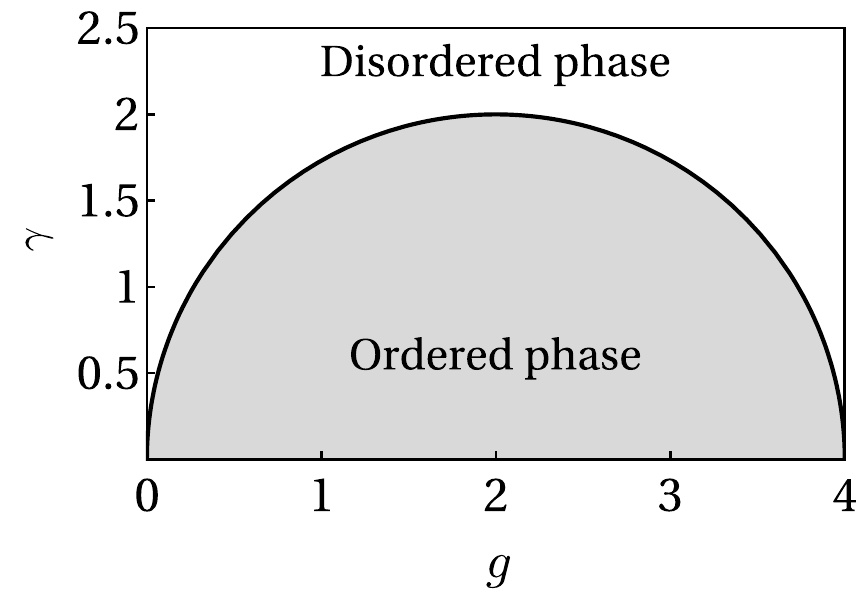}
 \caption{NESS phase diagram of the SQS belonging to the 
 dissipative dynamics given in Eq.~(\ref{eq:NESSL}).}
 \label{fig:NESSPD}
\end{figure}

We are now interested in how the non-equilibrium dissipative phase transition 
may affect the metrological protocol. 
 \textcolor{black}{Therefore, we study the influence of this transition on 
 the quantum Fisher information.
 As we have already mentioned in Sec.~\ref{ssec:ness}, the NESS is still 
 a coherent state whose displacement and squeezing parameters are
 altered with respect to the equilibrium scenario. Therefore, the general formula 
 for the quantum Fisher information, cf.
 Eq.~(\ref{quantumFisher}), remains valid in the NESS upon the correct replacement
 of the self-consistent parameters $\omega$ and $h$ from 
 Eq.~(\ref{eq:NESSsol}).}\footnote{Recall that the displacement parameter may be obtained from the 
 self-consistent parameters viz. $\alpha = h/\omega$.}
%

\begin{figure}[t]
 \centering
\scalebox{1}{\includegraphics{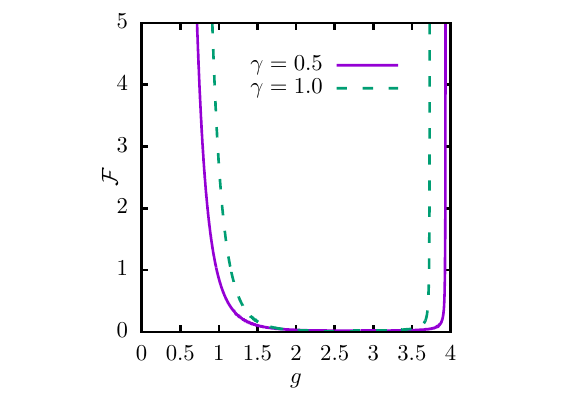} }
 \caption{ The quantum Fisher information $\mathcal{F}$ as a function of $g$ for the case with dissipation (out of equilibrium NESS case) in the ordered phase. 
 We see that the values for the $\mathcal{F}$ are lower than in equilibrium, see Fig. \ref{plots}. Nonetheless, the critical point, which is associated with arbitrarily large $\mathcal{F}$ is shifted to the left depending on $\gamma$.
 This may allow one to obtain larger precisions for other values of $g$ by using out of equilibrium scenarios.
 }\label{QDiss}
\end{figure}

In Fig. \ref{QDiss}, the quantum Fisher information $\mathcal{F}$ is plotted for two different values of the dissipation rate $\gamma$.
As discussed before, there are two phase transitions now, and therefore two critical points, as shown in Fig. \ref{fig:NESSPD}.
The first, on the right-hand side in Fig. 5, is a transition where increasing quantum fluctuations
destroy order. In turn, the second, on the left-hand side in Fig.~\ref{QDiss}, is a true non-equilibrium
transition where increasing quantum fluctuations induce order.
At these two critical points, the quantum Fisher information diverges, and therefore arbitrarily large precision may be obtained. 
For the critical value of $g$ for the transition on the right-hand side in Fig. \ref{QDiss}, as $\gamma$ increases, its value decreases as predicted in Fig. \ref{fig:NESSPD}.
In this way, the critical point can be shifted to the left by increasing $\gamma$.
By doing this, one is able to obtain larger precisions to estimate $g$ in this region, when compared to the equilibrium case  $\gamma=0$. 
To be more specific, one fixes a value of $g$ in that region, and evaluate the quantum Fisher information $\mathcal{F}$ with $\gamma=0$. One finds a finite value for $\mathcal{F}$, as depicted in the left panel of Fig. \ref{plots}. Then, by introducing dissipation, one can make the chosen value of $g$ arbitrarily close to a critical point, what makes the value of $\mathcal{F}$ increases as much as one wants. This shows that dissipation may
help in the estimation of the quantum parameter $g$ of the SQS.

\section{Summary}
\label{sec:discussion}

We investigate how collective quantum mean field systems may be useful for metrological protocols by using equilibrium and out of equilibrium states.
The SQS, a simple system which possesses a quantum phase transition between an ordered and a disordered phase, is used as a toy-model.
We focus on the estimation of the parameter driving the phase transition, the quantum parameter $g$.
As a figure of merit for the quantum metrological protocol, we use the Fisher information and quantum Fisher information, the latter being an upper bound for the first.
As a result, the quantum Fisher information is associated with the largest possible precision in the metrological protocol with quantum systems.

For equilibrium states, we find that the quantum Fisher information diverges at the critical point ($g=4$), meaning that one can obtain arbitrarily large precision as $g\rightarrow g_c = 4$.
This also happens in the classical limit $g\rightarrow 0$, and therefore the estimation of small values of $g$ can be done with increasingly large precision as $g$ decreases.
We also evaluate the Fisher information for different photon count measurements, corresponding to different values of $\Omega$.
We found some dependence on  $\Omega$ for a certain interval of $g$. Interestingly enough, different sets of projectors (different values of $\Omega$) lead to  the same precision in the estimation of $g$ as it gets arbitrarily close to the critical point ($g_c=4$).

We then studied the impact that a dissipative driving may have on the parameter estimation. Therefore,
we coupled the SQS to a heat bath at zero temperature that does not thermalize the eigenmode of the SQS. In this manner, 
the competition between the coherent and the dissipative drive create an ordered and an disordered phase out of equilibrium
with two distinct phase transitions due to a re-entrance phenomena in the NESS phase diagram of out of equilibrium states.
Arbitrarily large precisions are now observed at both these phase transition points.
Since the two critical points where the Fisher information diverges depend explicitly on the dissipation rate $\gamma$, 
dissipation may help the estimation of values of the quantum parameter $g$ in the ordered phase
when compared to the case of thermal equilibrium at $T=0$. 
 This is because, while at equilibrium the system may not be critical, 
we can move it to its NESS transition by dissipatively driving it, therefore obtaining a significant improvement 
in the parameter estimation. 
We also note that, since criticality is a resource for quantum metrology \cite{Zanardi}, the thermal regime may also be useful for parameter estimation in a certain range of temperature, as the only critical point is shifted to the left as the temperature increases, see Fig. \ref{fig:phases}.
As a result, it would be interesting to study this in detail elsewhere.
This effect is quite counter-intuitive, since one would expect a thermal bath 
to interfere with any measurement in a negative way. We thus conclude that it may be rather beneficial
to measure systems far away from equilibrium.

\section{Acknowledgement}
The authors acknowledge fruitful discussions with GT Landi.
SW is grateful to the quantum information group at UFABC for warm hospitality where this 
work was initiated. Moreover SW and FLS acknowledges funding by CAPES under CAPES/PrInt 
- process no. 88881.310346/2018-01.
SVM acknowledges Brazilian agency CAPES for financial support. FLS also
acknowledges partial support from of the Brazilian National
Institute of Science and Technology of Quantum Information
(CNPq INCT-IQ 465469/2014-0) and CNPq (Grant No. 302900/2017-9).


\appendix

\section{Quantum Fisher information}
\label{app:qfi}

\subsection{Infinitesimal Squeezing}

We shall evaluate this derivative using an infinitesimal squeezing transformation 
since this approach highlights the different contributions arising from displacement
and squeezing operations. Therefore, we proceed by formally writing the derivative of the SQS ground-state with respect to the parameter $g$ as difference quotient as
\begin{equation}
 \frac{d}{dg} \ket{0_{\text{\sc sqs}}(g)} 
= \lim_{\epsilon \to 0} \frac{1}{\epsilon}\bigg( \ket{0_{\text{\sc sqs}}(g+\epsilon)}-
\ket{0_{\text{\sc sqs}}(g)} \bigg)
\end{equation}
\begin{figure}[b]
 \centering
 \includegraphics[width=.3\textwidth]{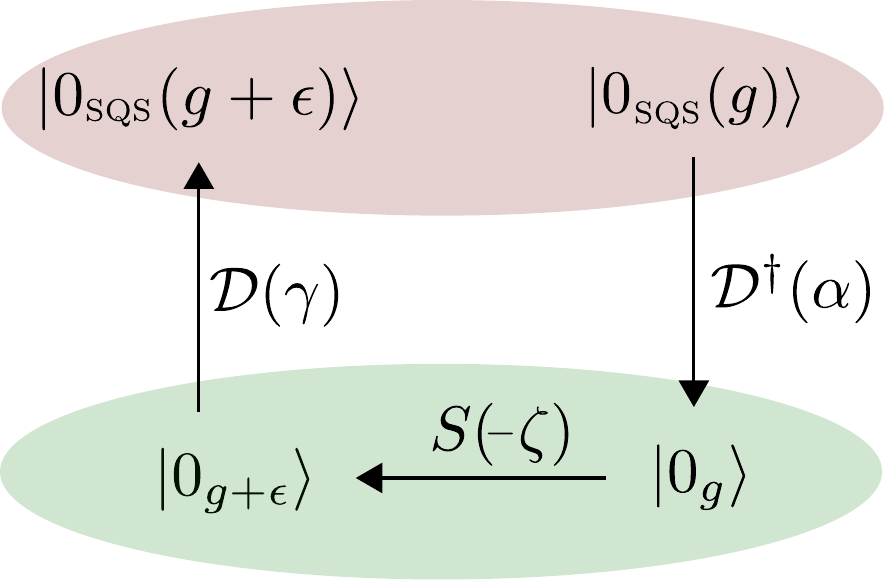}
 \caption{Connecting SQS ground states belonging to different quantum coupling
 parameters.}
 \label{fig:trafo}
\end{figure}
Fig.~\ref{fig:trafo} illustrates how the distinct ground-states are connected via a
sequence of squeezing and displacement operations.
In this manner the expression
\begin{align}
 \frac{d}{d g} \ket{0_{\text{\sc sqs}}(g)} 
&= \left[\frac{d \mathcal{D}(\alpha)}{d g}  S(\zeta)   + \mathcal{D}(\alpha)\frac{d S(\zeta)}{d g} \right] \ket{0_g}
\end{align}
is readily deduced.
The derivative of the displacement operator for a real-valued displacement $\alpha \in \mathbb{R}$
may be generally 
written as
\begin{align}
 \frac{d\mathcal{D}(\alpha)}{dg} 
&=f(g)\left( a^\dagger - a\right)\mathcal{D}(\alpha), \quad \text{with}
\label{eq:dD}\\
f(g) &= \frac{d\alpha}{d g} 
+\frac{\alpha(g)}{2g}
\frac{ \omega(g)-g\omega'(g)}{\omega(g)}
\end{align}
The derivative of the squeezing operator for real-valued squeezing parameters $\zeta
\in  \mathbb{R}$ is given by
\begin{equation}
 \frac{d S(-\zeta)}{dg} = -\frac{1}{2} \frac{d\zeta}{dg} \left( a^2
 -\left(a^\dagger\right)^2 \right) S(\zeta)
\end{equation}
In this way, the derivative state is cast in the compact form
\begin{align}
\partial_g \ket{0_{\text{\sc sqs}} (g)} &=
f(g)\ket{1_{\text{\sc sqs}} (g)}  +\frac{1/g-\omega' /\omega }{2^{3/2}}\ket{2_{\text{SQS}}(g)} 
\end{align}

As a double check of this formula, we derive it next using the position representation
of the harmonic oscillator.

\subsection{Position Eigenstate}
It is possible to avoid the infinitesimal squeezing and make use of the position representation
of coherent states in order to calculate the derivative. We shall briefly sketch this
calculation here.
In order to avoid the infinitesimal squeezing we shall work with the harmonic oscillator Fock
state viz
\begin{equation}
 \frac{d}{dg}\ket{0_{\text{SQS}}(g)} = \frac{d\mathcal{D}(\alpha)}{dg} \ket{0_g}
 +\mathcal{D}(\alpha) \frac{d}{dg}\ket{0_g}
\end{equation}
The first term is identical to the previous calculation, compare Eq.~(\ref{eq:dD}). For
the second term we may introduce a position representation viz
\begin{widetext}
\begin{align}
 \frac{d}{dg}\ket{0_{\text{SQS}}(g)} 
%
&= \frac{d\mathcal{D}(\alpha)}{dg} \ket{0_g}
 +\mathcal{D}(\alpha)\int_{-\infty}^\infty dx  \frac{d}{dg}\left[
 \left(\frac{\omega}{\pi g}\right)^{\frac{1}{4}} e^{-\frac{\omega}{2g}x^2}
 \right]\ket{x}\\
&= \frac{d\mathcal{D}(\alpha)}{dg} \ket{0_g}
 +\mathcal{D}(\alpha)\int_{-\infty}^\infty dx  \left[
 \frac{1}{4} \left(\frac{\omega}{\pi g}\right)^{-\frac{3}{4}}\frac{\omega' g 
 - \omega}{\pi g^2}-\left(\frac{\omega}{\pi g}\right)^{\frac{1}{4}}
 x^2 \frac{\omega'  g -  \omega}{2g^2} 
 \right]e^{-\frac{\omega}{2g}x^2}\ket{x}\\
 %
 %
&= \frac{d\mathcal{D}(\alpha)}{dg} \ket{0_g}
 +\mathcal{D}(\alpha)\frac{1/g-\omega' /\omega }{4^{3/4}}\ket{2_g}\\
&= f(g) \ket{1_{\text{SQS}}(g)}+
 \frac{1/g-\omega' /\omega }{2^{3/2}}\ket{2_{\text{SQS}}(g)} 
 \end{align}
\end{widetext}

\newpage \ \newpage

\bibliographystyle{apsrev4-1}
\bibliography{manuscript_v2}

\end{document}